# Lightweight Multilingual Software Analysis

Damian M. Lyons[1], Anne Marie Bogar[1] and David Baird[2]

[1]Department of Computer and Information Science, Fordham University, New York NY U.S.A.
[2]*Bloomberg L.P., New York NY U.S.A.*
{dlyons, abogar}@fordham.edu, dbaird16@bloomberg.net

**Abstract.** Developer preferences, language capabilities and the persistence of older languages contribute to the trend that large software codebases are often *multilingual* – that is, written in more than one computer language. While developers can leverage monolingual software development tools to build software components, companies are faced with the problem of managing the resultant large, multilingual codebases to address issues with security, efficiency, and quality metrics. The key challenge is to address the opaque nature of the language interoperability interface: one language calling procedures in a second (which may call a third, or even back to the first), resulting in a potentially tangled, inefficient and insecure codebase. An architecture is proposed for lightweight static analysis of large multilingual codebases – the MLSA architecture. Its modular and table-oriented structure addresses the open-ended nature of multiple languages and language interoperability APIs. We focus here as an application on the construction of call-graphs that capture both inter-language and intra-language calls. The algorithms for extracting multilingual call-graphs from codebases are presented, and several examples of multilingual software engineering analysis are discussed. The state of the implementation and testing of MLSA is presented, and the implications for future work are discussed.

# Introduction

Large software projects may typically have components written in different languages. Companies that have a large software codebase may face the issue of applying security, efficiency and quality metrics for a product spanning many languages [1] [2]. A developer or developer organization may choose one language for numerical computation and another for user interface implementation, or they may have inherited or be mandated to work with legacy code in one language while extending functionality with another. While there are many such drivers promoting multilingual codebases, they come with significant software engineering challenges. Although a software development environment might support multiple languages (e.g., *Eclipse IDEs*) it may leave the language boundaries, *language interoperability* [3], opaque. While it may be possible to automatically inspect individual language components of the codebase for software engineering metrics, it may be difficult or impossible to do this on a single accurate description of the complete multilingual codebase.

Our objective is to develop software engineering tools that address large multilingual codebases in a lightweight, open and extensible fashion. The MLSA (*MultiLingual Software Analysis*) architecture – a lightweight architecture concept for static analysis of multilingual software – is presented**.** There are several places where language boundaries raise software engineering challenges in multilingual software; in this paper we focus on one of those – language interoperability, where functions in one language call functions in another. One of the key tools and prerequisites for several kinds of software analysis is the call-graph. Because the call-graph is also where language boundaries directly meet for the interoperability domain, we will focus on the issues of generating multilingual call-graphs, and in particular, for C/C++, Python and Javascript interoperability examples. Several implementation examples are presented showing the benefit of multilingual call-graph analysis from the perspectives of efficiency and security. Our results are summarized, and future directions discussed in the final section of the paper.

## 2  Prior Literature

A multilingual codebase can arise for historical reasons. As languages rise and fall in popularity, their varying use results in a codebase that reflects developer or commercial historical preferences [4]. Libraries for numerical computation may have been constructed in FORTRAN, C and C++ for example, and front-end libraries may have been built in JavaScript. But there are also good software engineering reasons for choosing to develop different parts of a project in different languages: Toele [5] argues for increased developer productivity among other benefits.

Nonetheless, a multilingual codebase also poses significant software engineering challenges. Although multilingual code is common, development tools tend to be language specific, with some cross-platform functionality. As one example among many, *Checkmarx[1]* offers static analysis [6] for a wide range of languages, but individually. While each language codebase can be analyzed, the transitive closure of calls across language boundaries is harder to capture. The ways in which language boundaries are crossed in a multilingual code base include:

- Programs in one language that communicate via inter-process communication (IPC) with program in other languages.
- Programs in one language that call functions that have been defined in a second language (Interoperability)
- Files that contain functions in one language embedded in the syntax of a second, for example, SQL commands embedded in a Java program, or JavaScript scripts embedded in an HTML document.

Multilingual codebases present more challenges for developers. Mayer et al. [7] survey developers and tools, and reported that they find that 90% of their developers request help in developing multilingual code bases. The challenges faced include the follow:

---

[1] www.checkmarx.com

- Redundancy, e.g., procedures in several different language libraries for the same functionality, necessitating refactoring [8].
- Debugging complexity as languages interact with each other in unexpected ways [9].
- Security issues relating to what information is exposed when one language procedure is called from another [10] [11].

Recently several comprehensive reviews and surveys of the literature in this area have appeared [12] [13] [7]. The field has been active for over two decades (at least) and is very diverse in application and method. Zaigham et al. [12] report an analysis of over 3800 papers over the last 15 years in the area of multilingual source code analysis. However, they find that 23% of these papers were written in the last two years alone. One early proposed approach to handling the issues of multilingual systems is to use a versatile monolingual environment [14], but of course this is not too useful an approach for existing multilingual codebases. Synytskyy et al [15] addressed automatic software analysis of multilingual documents, e.g., dynamic web pages which contain HTML and scripts, Microsoft ASP that included server and client-side code in a single file, or Java/Cobol programs with embedded SQL statements. Island Grammars [16] are used to parse all the language components together in a single parse tree for analysis. Strien et al. [8] address cross-language program analysis and refactoring in their system X-DEVELOP, an IDE that includes meta-model information linking different language components by mapping between their abstract syntax trees (ASTs).

More recent approaches include the metalanguage Rascal [2], which provides tools with which program analysis algorithms can be written for different languages. Of course, this does not specifically address the problems that arise due to the language boundaries. Nguyen et al. [17] address cross-language program slicing – that is, determining the impact of a variable within a program. They focus on PHP web applications and cross-language data flows between HTML and PHP. Unlike the syntax-based approaches presented before, they use symbolic execution to generate and relate server and client-side data flows to support thin slicing. In [18] the same authors address the issue of constructing call-graphs for the same dynamic web page application. They make the point that much client side-code is generated from server-side code and hence both need to be analyzed in tandem. Again, symbolic execution is used to estimate client-side data including conditional structure, but in this case parsing is needed to find calls to build the call-graph.

The approach proposed in this paper different from the prior literature in several key aspects. First, while we present results here for multilingual call-graph analysis, we do so in the context of constructing a light-weight modular system for analyzing multilingual software. While the prior approach most similar to ours is that of Strien et al. [8], we will analyze monolingual ASTs and we leverage the existing interoperability API descriptions (such as, e.g., *python.h* for C programs calling Python) to map between these ASTs as a common meta-model. However, a key challenge we face in using interoperability API descriptions is that the name of the target (i.e., called) procedure is passed as an argument to the interoperability API call. Rather than using symbolic execution to determine this kind of run-time dependency (as in Nguyen et al. [17] [18]), we leverage an existing static analysis technique called Reaching-Definitions Analysis [19][2] to determine possible values for the called procedure. Our approach is more directed and more 'bare bones' – targeted primarily at the language interface and using little extra infrastructure.

In the next section, we will describe the architecture for *MLSA*, a set of lightweight modular tools for multilingual software analysis. There are many important software metrics and analyses for large software architectures [20]. In the subsequent section, we delve into one specific analysis: call-graph generation for multilingual codebases.

## 3    The MLSA Architecture

It is very likely that new computer languages will continue to be developed and used, while existing languages will also tend to persist [4], leading to an increasingly crowded and complex landscape of multilingual software development. Any solution that is narrowly focused on the existing state-of-the-art will find itself quickly outdated as new languages, or interoperability APIs, or language embeddings appear. For this reason, we propose the following design principles for a software architecture for *MultiLingual Software Analysis* - the MLSA Architecture.

---

[2] A points-to analysis would be a more comprehensive solution, but is strictly speaking, more than is required for the restricted problem as described later in this paper.

### 3.1 MLSA Design Principles

1. Lightweight: Computation is carried out by small programs, which we call *filters*, that communicate results with each other. A specific software analysis, such as multilingual call-graph generation, is structured as a pipeline of these programs. Other analysis may use some of the same filters.

2. Modular: Filter programs accept input and produce output in a standard form, to allow modules to be substituted or added with minimum collateral damage.

3. Open: MLSA uses open-source software for monolingual processing and for display, and it is open-source friendly to the addition of new filter programs and pipelines because of the previous two design principles (lightweight and modular).

4. Static Analysis: MLSA uses static analysis as its principle design tool. Static analysis is a broad analysis in that it looks at all potential executions before code is executed and does not require executing annotated software as does Dynamic Analysis. (But we come back to this in the last section.)

A generic example software analysis MLSA pipeline is illustrated in Figure 1. Pipelines are generally divided into three layers:

1. The initial filter programs consume a monolingual AST generated by an appropriate monolingual parser in the *monolingual layer* of the architecture. These ASTs are generated by existing open source tools such as Clang[3].

2. Because each language AST differs, the programs that consume the monolingual ASTs to process interoperability APIs must be also language specific; this happens in the *interoperability layer*.

3. In the final, *multilingual layer*, all the program data has been transformed to multilingual, and procedures in different languages can be related to one another for static analysis.

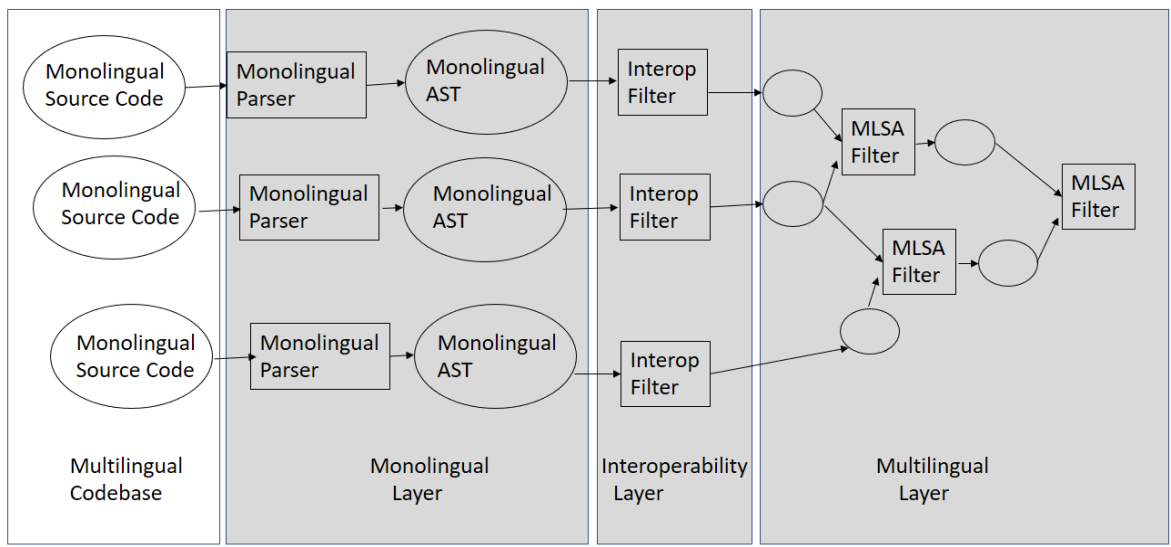

**Fig. 1.** Example MLSA software analysis showing filter programs (squares) and their standardized input and output (circles) operating on a multilingual codebase (unshaded right-hand side).

---

[3] clang.llvm.org

As an illustrative example of Fig. 1, let us consider a Reaching Definitions Analysis [19] (RDA) pipeline for a C program.
- A monolingual AST filter `CASN` inputs the C AST and generates an output data file of variable assignment locations. As a MLSA convention, all output data files are in a tabular format.
- A second filter `CRCF` generates the reverse control flow as a data file.
- A multilingual filter `RDA` takes the variable assignment location table and reverse control flow table as inputs, calculates reaching definitions for each variable, and writes those to a data file.

Each filter is a small standalone program, implementing a single analysis, illustrating the lightweight design principle. The only constraint on the program is on the format of its input and output. Filter programs can be written in different languages or can be shell scripts.

To illustrate the modularity design principle let us consider what is required to extend the RDA analysis to handle both Python and C software:

1. Adding extra languages just requires adding new monolingual filters for the language. For Python, we need to add `PASN` and `PRCF` filters, analogous to `CASN` and `CRCF`. Each filter just needs to read its monolingual AST and generate a tabular data file in the standard format associated with that filter.
2. Modifying analyses just requires reconfiguration of the analysis pipeline. If, for example, the filter program `CCON` extracts the locations of conditional statements (the condition in an IF or WHILE) and outputs this in the same tabular format as `CASN`, then merging the output of `CCON` with that of `CASN` and providing that table as input to `RDA` would perform RDA for the condition statements.

The open design principle is why we chose to use existing monolingual parsers rather than construct special parsers. Furthermore, a developer who wants to contribute new functionality to MLSA can add new filter programs or new pipelines of existing filters with little collateral effect.

### 3.2 Computational Efficiency

Although not one of the design principles, the MLSA architecture was also designed with computational efficiency in mind. The policy of dividing the analysis into pipelines has the effect of making parallelism and data-dependency explicit and this can be leveraged for computational efficiency. For relatively small multilingual codebases, a static analysis network (as in Figure 1) could be distributed among multiple cores of a multicore processor. The parallelism and dependency can be derived directly from the pipeline.

For a large codebase, in a realistic large company scenario with a widely-distributed set of code developers and contractors, a static analysis network might need to function continually (a daily basis for example) on cloud computing, regenerating and updating tabular data file components across the network in a parallel distributed fashion as guided by the pipeline description.

## 4 Multilingual Call-graph Analysis

Call-graph analysis (CGA) is a useful software engineering tool [21]. For multilingual code, the call-graph can be used to investigate the boundary line between languages – a boundary that is opaque in many tools. For example, a C program may call a Python procedure in addition to many C procedures. Consider that one such C procedure `OpenPort` exposes a security risk and needs to have its invocations pass a security review. Just looking at the call-graph of the C procedures, some of which may invoke the Python procedure, can give false confidence that it shows all the call sequences for the program. However, the Python code may itself call `OpenPort` or may call other procedures in that same C program that in turn call `OpenPort`.

This section details a specific analysis implemented on the MLSA architecture: creation of a multilingual call-graph. The implementation relies on monolingual AST filters for C/C++, Python and JavaScript to generate tabular data files containing monolingual call-graph information (monolingual layer). These are in turn processed by language boundary filters that characterize specific cross-language interoperability APIs, generating modified data files (interoperability layer). These are integrated into a single, multilingual call-graph (CG) at the multilingual layer. This pipeline is shown in Figure 2 for an example where a multilingual codebase with mutual calling between three different languages is resolved into a single multi-lingual CG.

Figure 2 shows three pipelines – one per language – that join for the last step, the construction of the multilingual CG. Each monolingual AST (*MonoIng AST* in Fig. 2) is processed to produce reverse control flow, assignments and a monolingual CG. Interoperability API calls appear in this monolingual CG as straight procedure calls. Recall that the control flow and assignment locations are necessary to carry out RDA to determine the values of parameters to interoperability calls; without knowledge of the values, it may not be possible to determine which cross-language procedure is being called (target) and hence construct the multilingual CG.

The interoperability filter (*Interop Filter)* is responsible for identifying interoperability API calls in the monolingual CG and using the results of the RDA analysis to resolve the call to its target, cross-language procedure if possible. It may not be possible for RDA to determine the target if the target was the result of a run-time calculation. RDA may also produce multiple potential values since it conservatively interprets the effects of conditional statements. We will come back to these two cases later in this section.

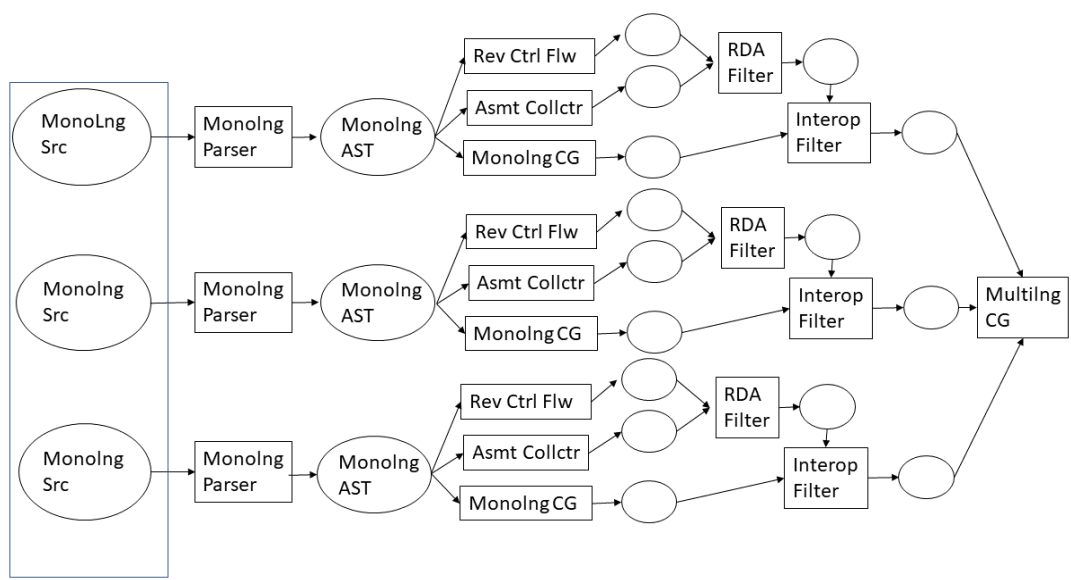

**Fig. 2.** Pipeline for generating multilingual call-graphs

The final step in the pipeline is the integration of the modified monolingual CGs – modified with the information about the language, names and arguments of the targets of any interoperability API calls – into a single multilingual CG. Our specific focus in this paper is on the interoperability filter program (Interop Filter) and on the integration of its results into a multilingual CG (*Multilng CG*). In the following sections, we formalize the operations of these programs.

### 4.1 MLSA CG Construction

Calling a cross-language procedure may be almost trivial (e.g., C with C++) or may involve a cross-language API as in the case of JNI (C with Java) or *Python.h* (C/C++ with Python) and others. A monolingual CG analysis for a program in $S_L$ in language L, which we write $CG(S_L)$, will yield leaf nodes in $CG(S_L)$ that are the cross-language API calls. We begin with some definitions.

**Definition 1**. Let us consider the program $S_L$ in language $L$ to be a set of $(\ell, b)$ basic blocks $b$ (elementary statements) with statement number or label $\ell$. The set $B$ of elementary statements includes a procedure call statement, and for $(\ell, b)$, $b \in PC \subseteq B$ a procedure call, we define:
- *target(b)* is the name of the called procedure
- *arg(b)*=$a_0,...,a_n$ are the arguments of the call
- *lab(b)* = $\ell$
- *procs($S_L$)* = { *target(b)* : $b \in PC$ }

**Definition 2.** We define a *call-graph* for the program $S_L$ as $CG(S_L)=(V_L, E_L)$ where:
- $V_L=\{(p, \ell, a)\}$, where
  - *p=target(b)* for some block $b$ is a procedure called in $S_L$,
  - $\ell$ =*lab(b)* is the statement number of the call, and
  - *a=arg(b)* is the argument list in the procedure call; and,
- $E_L \subseteq V_L^2$ links a node $v$ to node $u$ *iff* some execution of $p_v$ calls $p_u$ with arguments $a_u$.

**Definition 3.** We write a multilingual codebase D as a set of programs $\{ S_L^i : i \in 1...N_D, L \in W_D \}$ where $S_L^i$ is uniquely identified by $i$ and it is written in language $L$. There are $N_D$ programs in the codebase and $W_D$ lists all the languages used in the codebase.

The construction of a multilingual CG from a set of monolingual programs with some mutual operability calls can be described as a two-step process. As the first step, an interoperability filter $IF_{L,M}$ takes $CG(S_L)$ as input and, using the description of the interoperability APIs in language $L$ for each language $L' \in M$, generates a multilingual call-graph $CG_M$ in which the interoperability calls in language $L$ with target in language $L'$ have been replaced with their target calls in $L'$ as specified by the parameters in the operability API calls. These target calls are still leaf nodes however, and are annotated with the name of the target language $L'$. For example, consider a C program *S.c* that calls some Python procedures defined in *S.py*. The Python call-graph filter generates a call-graph $CG_P(S.py)$, and the C call-graph filter similarly generates $CG_C(S.c)$. The interoperability filter $IF_{C,\{P\}}$ acts on $CG_C(S.c)$ and replaces the interoperability calls to Python procedures with the names and parameters of the target procedures, producing $CG_M(S.c)$.

The second step in generating the multilingual CG is to extend these leaf nodes in $CG_M$ by the definitions of the target procedures. The multilingual CG filter *MCG* takes $CG_M$ as input and for each leaf node $v$ in language $L'$, it determines if there exists an $S_{L'}^i$ in the codebase $D$ that defines $v$. If so, it extends $CG_M$ at node $v$ with the subgraph of $CG(S_{L'}^i)$. Continuing with the example from the previous paragraph, MCG takes $CG_M(S.c)$ as input and, locating the definitions of each target procedure $v$ in $CG_P(S.py)$, it extends $CG_M(S.c)$ with the subgraph of the definition of $v$ from $CG_P(S.py)$.

Before we present detailed algorithms for these two steps, it is necessary to note that here are many different language interoperability APIs for executing code or calling procedures from one language in another. And more are likely being added all the time. Based on our study of the existing interoperability APIs, it's possible to cluster how these support procedure calls into three groups: *Anonymous, File-based* and *Procedure-based*. Each of these will be addressed in turn, showing how they can be leveraged to generate a multilingual call-graph from separate monolingual call-graphs and rules for processing the interoperability API.

## 4.2 Anonymous Interoperability

The most typical form of interoperability API allows a string containing code in one language to be executed in a second. Examples of this include:

- *Python.h*: *PyRun_SimpleString()* - executes a string of Python commands in C/C++.

- *Stdlib.h*: *system()* - executes a Linux shell command from C/C++. Variants exist in many languages, such as Python's *os.system()*.

- PyV8: *PyV8.JSContext.eval()* – executes a JavaScript string from Python.

- Emscripten: *emscripten_run_script()* – executes a JavaScript string from C/C++.

Because these APIs allow an unlabeled block of code to be executed from a single procedure call in the program, we refer to them as *anonymous multilingual procedures*. They pose many issues for software engineering because the code string may be opaque to code analysis: the code may be created at run time, making it difficult to determine a-priori if the code abides by software engineering criteria design to promote correct, maintainable, efficient and safe code.

The approach proposed here is to add an edge to the CG for each possible value of the argument string that can be statically identified (e.g., a string literal). The edge leads to a special node labeled *Anonymous*. In the case that at least one possible value of the argument string cannot be identified, that node is labeled *Anonymous-Dynamic*. Although such multilingual code may in fact be correct and safe, it is certainly worthwhile from a software engineering perspective to label it clearly for more thorough inspection. Before we present the code, we need to clarify how our reaching-definitions analysis is defined.

**Definition 4**. The reaching definitions analysis filter is defined as follows: $RDA(p, X, \ell) = \{(x, \ell') : x \in X\}$ to be the statement number $\ell'$ of the last assignment in procedure $p$ for each variable $x$.

The pseudo code for the interoperability filter to process anonymous calls such as `PyRun_SimpleFile` is as follows (using Definitions 1 through 4).

```
Algorithm 1: InteropFilter( CG(SL) )
1.  CG(SL) = (V,E)
2.  For each v = (pv, ℓv, αv) ∈ V with (u, v) ∈ E    // node and parent
3.    If pv = PyRun_SimpleString                     // or equivalent
4.      For (a0,ℓ') ∈ RDA(pu,{a0},ℓ), where αv=a0,…,an // definition of a0
5.        Calculate y ← Eval(a0, ℓ')                 // static evaluation
6.        If y=∅ pv ←"Anonymous-Dynamic"             // no static value
7.        Elseif |y|=1                               // only 1 static value
8.          pv ←"Anonymous", a0 ← y, a* ← L'         // a0 has code string
9.          Else For each x ∈ y                      // all possible values
10.           Copy v into v', add (u,v') to E        // copy this branch
11.           pv' ←"Anonymous", a0 ← y, a* ← L'
```

The RDA analysis determines the statement number $\ell'$ that the first argument to the interoperability API call was last assigned. The *Eval* function determines if the value can be statically evaluated. Not all values can, of course. So, in the case that it is not possible, this is marked using $\varnothing$. It is also possible for the first argument to have multiple values. In that case, branches are added to the call-graph for each value. Notice that when the value of the first argument is known, it is added to the graph. It is also necessary for the modified nodes to be labelled with the target language; This is accomplished here by adding an extra argument $a_*$ with the target language name, L' – to label the node for step two processing.

### 4.3 File-Based Interoperability

Another common form of interoperability API allows a filename for a program written in one language to be executed from a second language. The program will at least contain some statements and it may call other procedures. The objective is to add a node to the CG that represents calling the main program in the file as a procedure, and to add the procedure calls by this node as edges in the CG.

Examples of this form of cross-language API include:
- *Python.h*: *PyRun_SimpleFile(Filename)* – will execute the python code in Filename from a C/C++ program.
- PyV8: *PyV8.JSContext.eval(jsfile.read)* – executes the contents of a JavaScript file *jsfile* from Python.
- JQuery: *JQuery.ajax(url: "pyfile.py")* – executes a Python file from JavaScript.

If the possible values for the filename can be statically determined, then the code to be executed is directly visible to code analysis. It is reasonable to assume that a CG can be generated for the file, and the issue is just to stitch this in to CG for the program in the calling language. However, if the filename is unknown, then we have the same opacity issues as we had for anonymous multilingual procedures and thus this needs to be flagged for inspection.

The approach proposed here is to add an edge to the CG for each possible value of the filename, and to attach the call subgraph for the file at that node. The pseudocode to accomplish this is basically the same as that in Algorithm 1 with the following changes:

1. The file-based interoperability API call is tested-for in line 3.
2. The name of the file, if it is statically available, is added to the modified call-graph $CG_M$ at lines 8 and 11, and the arguments to the procedure are modified to omit that first argument.
3. The node is labeled "FileBased-Dynamic" at line 6

```
Algorithm 2: InteropFilter( CG(S_L) )
1.  CG(S_L) = (V,E)
2.  For each v = (p_v, ℓ_v, α_v) ∈ V with (u, v) ∈ E     // node and parent
3.    If p_v = PyRun_SimpleFile                          // or equivalent
4.      For (a_0,ℓ') ∈ RDA(p_u,{a_0},ℓ), where α=a_0,…,a_n  // definition of a_0
5.        Calculate y ← Eval(a_0, ℓ')                    // static evaluation
6.        If y=∅ p_v ←"FileBased-Dynamic"                // no static value
7.        Elseif |y|=1                                   // only 1 static value
8.          p_v ← y, a←a_1,…,a_n, a* ← L'                // y is target
9.        Else For each x ∈ y                            // all possible targets
10.         Copy v into v', add (u,v') to E              // copy this branch
11.         p_v' ← y, a←a_1,…,a_n, a* ← L'
```

### 4.4 Procedure-based Interoperability

The most sophisticated form of interoperability API call is where the procedures from a file written in one language are read into a context object in a second language, and then can each be called from the context object with procedure parameters. Examples of this include the follow:

- *Python.h*: *PyObject_CallObject(pFunc, pArgs)* – when *PyImport_Import()* is used to load a Python module, individual functions from the module can then be called with *pFunc*.
- JSAPI: *JS_CallFunctionName(cx,gl, fn, args, &ret)* – calls JavaScript function *fn* from a C program with arguments *args* and return *ret*.
- Python-Bond: *js.call( fname, args)* – calls JavaScript function *fname* with arguments *args* in Python.

This multiple step interoperability API means that the filename with source code is presented first to (one or more) procedure call(s) to create the multilingual context. If that argument can be statically evaluated, then a call-graph can be generated for all the procedures that can be called. If at least one of the possible values of the filename is unknown, then any call to a procedure from this context is labeled in the CG as a Procedure-Based Dynamic call to flag it for more detailed code review. The pseudocode to accomplish this is basically the same as that in Algorithm 2 with the following changes:

1. The procedure-based interoperability API call is tested for in line 3 (e.g., *PyObject_CallObject)*
2. The name of the procedure, if it is statically available, is added to the modified call-graph $CG_M$ at lines 8 and 11, and the arguments to the procedure are modified to omit the first argument.
3. The node is labelled "ProcBased-Dynamic" at line 6

This algorithm does not include several steps typically associated with procedure-based interoperability calls. Continuing our *python.h* example:

1. The first argument to *PyObject_CallObject* is a function pointer whose value is set by a call to *PyObject_GetAttrString()* with arguments that include the imported context and the name of the function to call.

2. The argument list is composed by *PyTuple_SetItem* and therefore not immediately in the same form as C/C++ procedure argument lists.

Thus, setting the name in lines 8 and 11 will typically require making a call to find the procedure name associated with the function pointer, and modifying the arguments on these same two lines may require unpacking them.

### 4.5 Completing the Multilingual Call-Graph

The final step is to extend each modified leaf node in $CG_M$ with the definition of the procedure from the target language CG. The multilingual call-graph $CG_M$ is constructed as follows:

```
Algorithm 3: MFG( CG_M )
1. CG_M= (V,E)
2. For each leaf v ∈ V with a* = L'
3.   If there exists S_{L'}^i in D with p_v ∈ procs(S_{L'}^i)
4.     CG_p ← subgraph of CG(S_{L'}^i) rooted at p_v
5.     Merge CG_M and CG_p
```

## 5 Implementation

In this section, we present several example analyses of multilingual CGs using MLSA. The current implementation covers C, Python and JavaScript languages. At the monolingual layer, the AST for C programs is generated by *Clang-Check*, those for JavaScript by *SpiderMonkey*, and a Python library *ast.py* used to generate the Python AST. At the interoperability layer, filter programs for *Python.h*, *PyV8* and *JQuery* support limited interoperability between the three languages. MLSA stores the data exchanged between filter programs in a tabular (CSV) form.

### 5.1 Multilingual Example

The graph in Figure 3 depicts a multilingual program in which a C++ program calls a Python program, and that Python program then calls a JavaScript Program. The C/C++ programs are represented by oval nodes, the Python programs by rectangular nodes, and the JavaScript programs by hexagonal nodes.

In our implementation, the MLSA programs can determine cross-language calls based on a set of interoperability APIs, mainly *Python.h*'s *PyRun_SimpleFile* for calling a Python program from a C/C++ program, *PyV8*'s *eval*() function for calling a JavaScript program from a Python program, and *JQuery's* ajax function for calling a Python program from a JavaScript program The MLSA pipeline contains separate programs for analyzing the AST of each program based on the language in which the program was written. These programs parse the AST to recover all the function calls, their arguments, and their scope (where they are found in the program – inside of a function or the main body of the program). The end-product of these computations is a CSV file for each program. The CSV files are processed by filters for the interoperability APIs. Those resulting CSV files are combined into one, multilingual CSV file, which is then used to create a tree of all the function calls in the system. The graph in Figure 3 is a rendering of a DOT file generated from the multilingual CSV by *GraphViz*[4].

---

[4] https://graphviz.gitlab.io

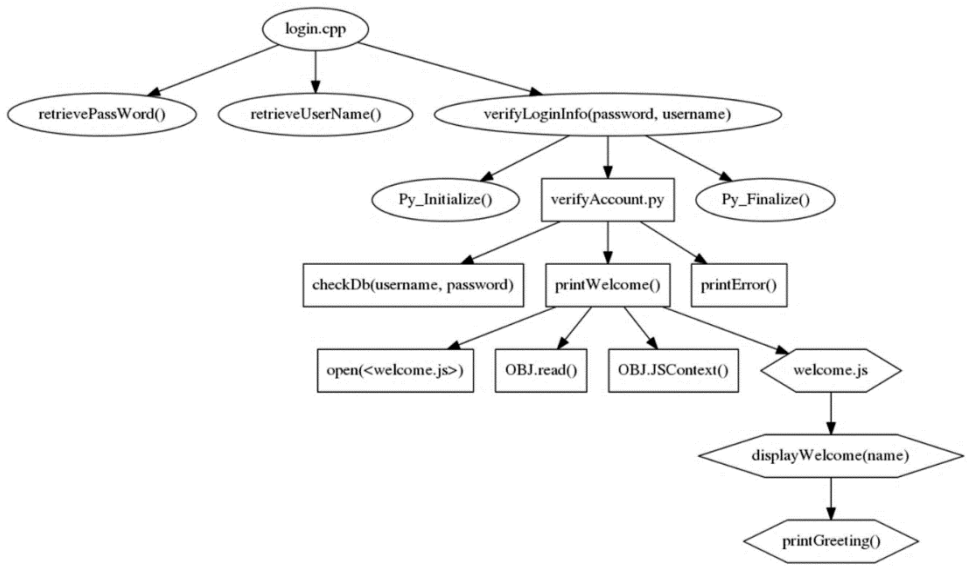

**Fig. 3.** Call-graph for a C program (main) that calls C procedures (ovals), Python procedures (rectangles) and JavaScript procedures.

### 5.2 Dynamic Call Example

Any multilingual calls that are not resolvable before run-time pose software engineering risks and should be flagged for review. Figure 4 depicts a potential security issue in which a Python program *verifyAccount.py* calls a JavaScript program, but the identity of the JavaScript program cannot be determined before run time.

This occurs in this case because the argument to the *js.call()* procedure is an expression. The value of the expression is computed as unknown in Algorithm 1 after the reaching definitions analysis. The call is still placed into the call-graph but flagged for inspection; visually indicated here with the double dotted outline.

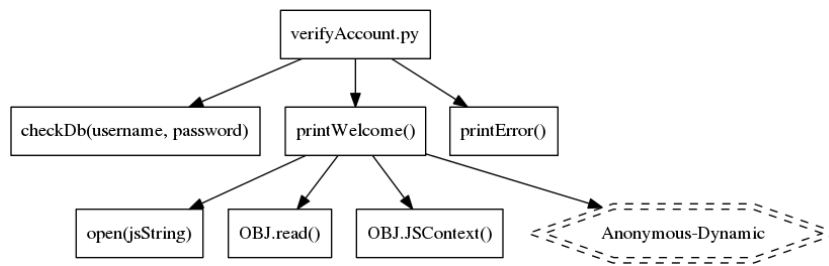

**Fig. 4.** Statically unresolved call from Python to JavaScript

### 5.3 Hidden Circularity Example

One of the main challenges we identified with a multilingual codebase is hidden circularity. For example, the C/C++ code may be validated free of unsafe C/C++ calls. But there may be interoperability API calls, which on their own seem safe, but when their CGs are inspected they result in unsafe calls back to the original C/C++ code.

The graph in Figure 5 is an example of such a circular system, in which the Python program *veriftyAccount.py* calls the JavaScript program *welcome.js*, and then the same JavaScript program *welcome.js* calls the Python program *verifyAccount.py*. The circularity of the system is flagged visually on the graph in Figure 4 by the node outlined with double dotted perimeter. The call may be fine, but because it is hidden from regular software engineering inspection behind the interoperability API, it requires another look.

The graph also shows a standard recursive call cycle in *displayWelcome()*. On the bottom right, the graph shows this recursive call as a node with dotted perimeter. Because such recursion is not problematic, it's not flagged for more inspection.

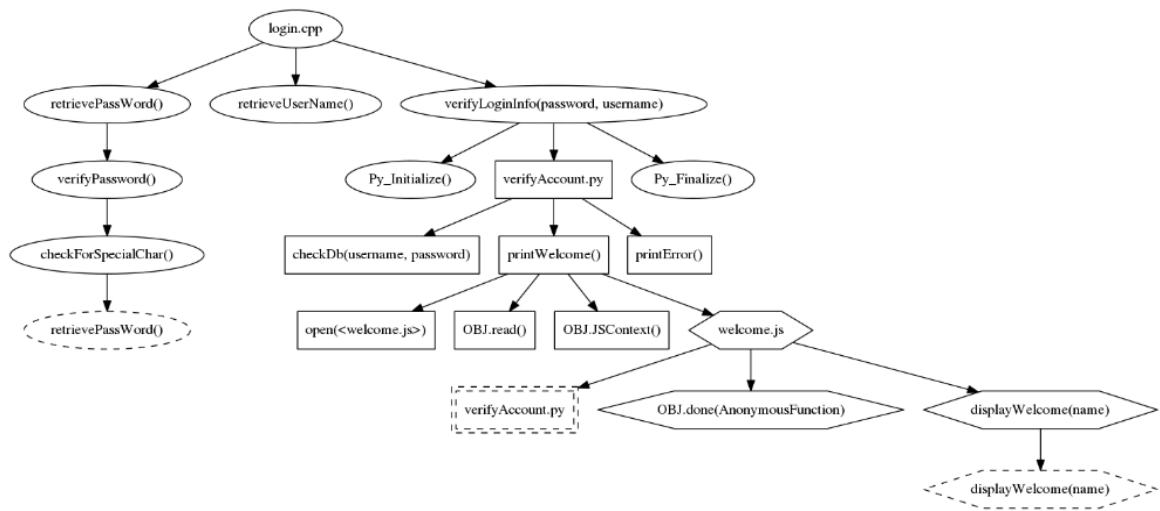

**Fig. 5.** Multilingual Call-graph with Circularity

## 7  Conclusion

This paper has addressed the issues faced by companies that must manage software architectures and libraries in different languages to enforce security, efficiency, and quality metrics. Because many existing software engineering tools are monolingual, even though multilingual code is widespread, issues relating to the language boundaries may get overlooked.

We have here motivated and proposed an architecture for analyzing multilingual software (MLSA) based on four design principles: Lightweight, Open, Modular and Static Analysis. Analyses are written in MLSA as pipelines of small filter programs operating on tabular data. We characterized three ways in which multilingual interactions can occur: through IPC, through interoperability and through embedding. We have chosen to focus here on interoperability and consequently picked an example application of generating multilingual call-graphs. Our objectives include analyzing all forms of multilingual interaction and we consider interoperability as a first step. One reason we chose interoperability is that solving this form of multilingual interaction has the additional challenge of estimating the values of interoperability API calls. We employed a static analysis (reaching definitions [19]) approach to address this. We also chose to focus on the non-OO subset of our target languages for simplicity. However, as a direction of future work, using a points-to analysis (which determines pointer values statically) rather than a reaching definitions analysis would support extending the call-graph analysis to objects as well improving the static evaluation of argument values.

Future work will also extend beyond interoperability and also look at embedded multilingual issues such as those that occur in Dynamic web pages. Simulated execution has previously been used [18] to address estimating parameter values for that application.

We focused on one specific problem, the generation of multilingual call-graphs and developed a detailed system diagram and approach for this. The C/Python interface is used as an example throughout, but the concepts presented are general, and equivalent variants in other interoperability APIS are listed. We partitioned the functions available in interoperability APIs into three general classes: Anonymous calls, File-based calls, and Procedure-based calls and we present a detailed description of the MLSA interoperability filters for each of these. Finally, we presented examples of the use of the MLSA multilingual call-graph filter in viewing multilingual files and detecting some software engineering issues in multilingual software.

We are currently in the process of evaluating the MLSA multilingual call-graph filter on readily available public codebases to measure its performance against existing monolingual call-graph generators and its precision in identifying language boundaries. An initial study of its performance on 70 C/C++/Python programs downloaded from the internet show that it worked without any issues on half of the programs and produced results, though with some missing information usually related to the OO constraint, on the other half. All multilingual language boundaries were however correctly identified. The call-graphs generated by MLSA are similar in representation to those generated by the Eclipse IDE, and future work will include a more detailed comparison of the MLSA call-graph filters with other available tools.

A number of other areas for future development have been identified. *Datalog*, a database query language that is a subset of Prolog, has been adopted for program analysis by some researchers [22] and there has been recent research in performance improvements [23]. Advantages to its use include the ease and conciseness of expressing program analysis rules in Datalog as opposed to, for example, C++ or Java, as well as the ease with which the results of separate analysis can be combined, due to the uniform representation. MLSA already stores the data exchanged between filter programs in a tabular (CSV) form that can be readily transformed to Datalog predicates to build the extensional databases that are the input for the filters.

One of the design principles for MLSA is the use of static analysis, chosen because it allows a broad range of operating conditions to be evaluated. This is as compared to dynamic analysis, which annotates code and evaluates execution traces for single runs or statistics for sets of runs. However, in prior work [24] [25], static analysis techniques have been extended with Dynamic Bayesian Networks [26] to produce probabilities associated with variable values. Integrating this with the RDA (or a points-to) filter in MLSA would mean that the results generated, while still sound, would be less conservative and more reflective of real run-time operation. This is a longer-term objective.

## Acknowledgements

The authors acknowledge the contributions of Bruno Vieira, Sunand Raghapathi and Nicholas Estelami in developing MLSA tools. The authors are partially supported by grant DL-47359-15016 from Bloomberg L.P.